\documentclass[twocolumn,prb,showpacs,preprintnumbers,amsmath,amssymb]{revtex4}

\usepackage{graphicx}
\usepackage{dcolumn}
\usepackage{bm}

\begin{document}

\title{Existence of a novel metallic ferromagnetic phase
in models for undoped manganites}

\author{Takashi Hotta}

\affiliation{Advanced Science Research Center,
Japan Atomic Energy Research Institute,
Tokai, Ibaraki 319-1195, Japan}

\date{\today}

\begin{abstract}
The existence of a novel metal-insulator transition in the ferromagnetic state
of models for undoped manganites is here discussed using numerical techniques
applied to the $e_{\rm g}$-orbital degenerate Hubbard model tightly coupled
with Jahn-Teller distortions.
The ground-state phase diagram is presented in the plane defined by the
electron-phonon coupling $\lambda$ and Coulomb interaction $u$.
In contrast to the standard one-band Hubbard model for cuprates,
the metallic phase is found to exist for finite values of
both $\lambda$ and $u$ in the present $e_{\rm g}$-orbital Hubbard model
even at half-filling, due to the Fermi-surface topology which is incompatible
with the staggered orbital ordering concomitant to the insulating phase.
Based on the present results, a possible scenario for
Colossal Magneto-Resistive effect is discussed in undoped manganites.
\end{abstract}



\maketitle


One of the most interesting issues in recent research activities in
condensed matter physics has been the elucidation of the mechanism of
Colossal Magneto-Resistance (CMR) phenomena in manganites.
\cite{Tokura,Dagotto1,Dagotto2}
Recently, it has been widely recognized that the CMR effect appears when
the ground-state of manganites can be converted from insulating
to ferromagnetic (FM) metallic by a small magnetic field.
The stability of the FM metallic phase in manganites
is understood due to the double-exchange (DE) mechanism,
based on the strong Hund's rule  coupling between mobile $e_{\rm g}$
electrons and localized $t_{\rm 2g}$ spins.
On the other hand, the insulating phase in manganites occurs
due to the coupling between degenerate $e_{\rm g}$ electrons and
Jahn-Teller (JT) distortions of MnO$_6$ octahedra,
leading to various types of charge and/or orbital ordering,
as observed experimentally.\cite{Dagotto1}

In undoped RMnO$_3$ (R=rare earth lanthanide ions),
the parent compound of CMR manganites for most R-ions,
the A-type antiferromagnetic (AFM) insulating phase appears
with the C-type ordering
of the $(3x^2$$-$$r^2)$- and $(3y^2$$-$$r^2)$-orbitals.\cite{Hotta1}
Quite recently, for R=Ho, a novel AFM phase called the ``E-type''
spin structure phase has been reported as the ground-state,\cite{Munoz,Kimura}
but this state is also insulating with the same orbital ordering
as that of the A-AFM phase.\cite{Hotta2}
In general, the metallic phase in manganites has been widely considered to
appear only after holes are doped into 
such insulating undoped materials.
Namely, by substituting La by alkaline earth ions such as Sr and Ca,
holes are doped into the $e_{\rm g}$-electron band and, due to the 
DE mechanism, the FM metallic phase is generated.
Most of the discussion in manganites has been concentrated on
the phases induced by hole doping into the A- or E-AFM state
with different values in the bandwidth of carriers.
In this framework, it is implicitly assumed
that the undoped system is always $insulating$.

However, in the theoretical phase diagram for undoped manganites,
which has been obtained by the numerical analysis of the DE model coupled
with the JT distortions,
a FM metallic phase has been observed for weak electron-lattice
coupling region.\cite{Hotta1,Hotta2,Yunoki}
This phase did not receive much attention since the focus of previous papers
was on the insulating phases.
In fact, while it is an interesting possibility to have a metallic phase
for RMnO$_3$, its existence has not yet been confirmed even from
the theoretical viewpoint,
since the previous calculations were carried out in small-size clusters
and in addition the Coulomb interaction was not included explicitly.
If the FM metallic phase is confirmed to exist adjacent
to the A- or E-AFM insulating state in the undoped limit,
{\it the CMR effect could in principle occur even without hole doping}.
Here, recall that at the heart of the CMR phenomena is the
two-phase competition between FM metallic and
insulating phases.\cite{Dagotto1,Dagotto2}
The insulating phase in undoped manganites has been
widely recognized both from experimental and theoretical investigations,
while the existence of the FM metallic phase has not been explored,
in spite of its potential importance.

In this paper, it is attempted to show the existence of the FM metallic phase
in undoped manganites.
For this purpose, the spinless $e_{\rm g}$-orbital Hubbard model tightly
coupled to JT distortions is analyzed by using numerical techniques
such as relaxation method for distortions and Lanczos algorithm for
exact diagonalization.
For small electron-phonon coupling or Coulomb interaction,
an orbital disordered phase specified with metallic characteristics
is observed in two and three dimensions,
while in one dimension, the orbital-ordered insulating phase
is obtained except for the non-interacting case.
In the phase diagram depicted in the plane spanned by the 
electron-phonon coupling and Coulomb interaction,
the metallic phase is found to exist in a wide range
of relevant parameters.
Then, it is concluded here that the metallic phase should exist
even in undoped manganites, and its experimental confirmation by
suitable chemical substitutions for RMnO$_3$ should represent
a new challenge to experimentalists of Mn oxides.


Models for manganites include five important ingredients
such as the kinetic term for the $e_{\rm g}$ electrons, the
Hund coupling between mobile $e_{\rm g}$ electrons and
localized $t_{\rm 2g}$ spins, the
electron-lattice coupling between $e_{\rm g}$ electrons
and distortions of the MnO$_6$ octahedra, the
Coulomb interaction among $e_{\rm g}$ electrons,
and the AFM coupling between neighboring $t_{\rm 2g}$ spins.
\cite{Hotta3}
Fortunately, for the purposes of this paper 
it is not necessary to consider all these interactions.
Namely, since the FM phase is our focus here,
$e_{\rm g}$-electron spins can be assumed to be perfectly polarized from
the outset due to the strong Hund coupling
between $e_{\rm g}$ electrons and $t_{\rm 2g}$ spins.
Then, in the present work a spinless model including
only charge and orbital degrees of freedom can be used.


The Hamiltonian $H$ studied here is, thus, given by the combination
of three terms: $H_{\rm kin}$, $H_{\rm el-ph}$, and $H_{\rm el-el}$.
The first term indicates the hopping motion of $e_{\rm g}$ electrons,
written as
\begin{equation}
  H_{\rm kin} = -\sum_{{\bf ia}\tau\tau'}
  t^{\bf a}_{\tau \tau'} d_{{\bf i}\tau}^{\dag}d_{{\bf i+a} \tau'},
\end{equation}
where $d_{{\bf i}{\rm a}}$ ($d_{{\bf i}{\rm b}}$) annihilates
an $e_{\rm g}$-electron in the $d_{x^2-y^2}$ ($d_{3z^2-r^2}$) orbital
at site ${\bf i}$, ${\bf a}$ is the vector connecting nearest-neighbor
sites,
and $t^{\bf a}_{\tau \tau'}$ is the hopping amplitude between
$\tau$- and $\tau'$-orbitals along the ${\bf a}$-direction,
expressed as
$t^{\bf x}_{\rm aa}$=$-\sqrt{3}t^{\bf x}_{\rm ab}$=
$-\sqrt{3}t^{\bf x}_{\rm ba}$=$3t^{\bf x}_{\rm bb}$=$3t/4$
for ${\bf a}$=${\bf x}$,
$t^{\bf y}_{\rm aa}$=$\sqrt{3}t^{\bf y}_{\rm ab}$=
$\sqrt{3}t^{\bf y}_{\rm ba}$=$3t^{\bf y}_{\rm bb}$=$3t/4$
for ${\bf a}$=${\bf y}$,
and
$t^{\bf z}_{\rm bb}$=$t$ with
$t^{\bf z}_{\rm aa}$=$t^{\bf z}_{\rm ab}$=$t^{\bf z}_{\rm ba}$=0
for ${\bf a}$=${\bf z}$, respectively.

The second term $H_{\rm el-ph}$ indicates the coupling of $e_{\rm g}$
electrons with distortions of MnO$_6$ octahedra, given by
\begin{eqnarray}
  H_{\rm el-ph} &=& \lambda \sum_{\bf i}
  (Q_{1{\bf i}}n_{\bf i}+Q_{2{\bf i}}\tau_{{\rm x}{\bf i}} 
  +Q_{3{\bf i}}\tau_{{\rm z}{\bf i}}) \nonumber \\
  &+& (1/2) \sum_{\bf i}
  (\beta Q_{1{\bf i}}^2+Q_{2{\bf i}}^2+Q_{3{\bf i}}^2),
\end{eqnarray}
where $\lambda$ is the dimensionless electron-phonon coupling constant,
$Q_{1{\bf i}}$ is the breathing-mode distortion, and
$Q_{2{\bf i}}$ and $Q_{3{\bf i}}$ are, respectively, the
$(x^2$$-$$y^2)$- and $(3z^2$$-$$r^2)$-type JT-mode distortions,
$n_{{\bf i}}$=
$d_{{\bf i} {\rm a}}^{\dag}d_{{\bf i}{\rm a}}$+
$d_{{\bf i} {\rm b}}^{\dag}d_{{\bf i}{\rm b}}$,
$\tau_{{\rm x}{\bf i}}$=
$d_{{\bf i}{\rm a}}^{\dag}d_{{\bf i}{\rm b}}$+
$d_{{\bf i}{\rm b}}^{\dag}d_{{\bf i}{\rm a}}$,
and
$\tau_{{\rm z}{\bf i}}$=
$d_{{\bf i} {\rm a}}^{\dag}d_{{\bf i}{\rm a}}$$-$
$d_{{\bf i} {\rm b}}^{\dag}d_{{\bf i}{\rm b}}$.
The second term in $H_{\rm el-ph}$ is the usual quadratic potential
for distortions and $\beta$ is the ratio between the
spring constants for breathing and JT distortion mode distortions.
In this paper, the distortions are treated adiabatically.\cite{note}

The third term $H_{\rm el-el}$ includes the Coulomb interactions, 
and it is given by
\begin{equation}
   H_{\rm el-el}=u \sum_{\bf i} n_{{\bf i}{\rm a}} n_{{\bf i}{\rm b}},
\end{equation}
where $n_{{\bf i}\tau}$=$d_{{\bf i}\tau}^{\dag}d_{{\bf i} \tau}$ and
$u$ is the on-site Coulomb interaction.
Note that $u$=$U'$$-$$J$, where $U'$ in the inter-orbital
Coulomb interaction and $J$ is the Hund's rule coupling
in the standard notation for the multi-orbital model.
In order to investigate the possibility of  metal-insulator
transition in $H$, the model can be analyzed from many perspectives,
due to the combination of these three terms.


Let us start our discussion considering  the JT model, defined as
$H_{\rm JT}$=$H_{\rm kin}$+$H_{\rm el-ph}$, 
obtained by dropping $H_{\rm el-el}$.
Since in this paper undoped manganites with one $e_{\rm g}$ electron
per site are analyzed, it is clear that the oxygen ions that are
shared by adjacent MnO$_6$ octahedra are all active,
and the concomitant distortions are $not$
independent from site to site.
As a consequence, the cooperative effect is crucially important in this case.
To consider this issue carefully, 
the simplest way is to optimize directly the
displacement of shared oxygen ions by using
relaxation techniques.\cite{Hotta1}
In practice, considering sites ${\bf i}$ and ${\bf i+a}$, the oxygen
in between is only allowed to move along the ${\bf a}$-axis 
(i.e. buckling and rotations are simply neglected).
Note that in one and two dimensions, some of the oxygen ions
are not shared by neighboring octahedra.
In this paper, for simplicity, the
positions of such unshared oxygen ions are assumed to be fixed,
since in the real materials 
those ions are not allowed to move freely
due to the effect of compensating cations.
In other words, the one-dimensional (1D) chain or the two-dimensional
(2D) plane are not isolated, but they are 
embedded in a three-dimensional (3D) environment.

\begin{figure}[t]
\includegraphics[width=1.0\linewidth]{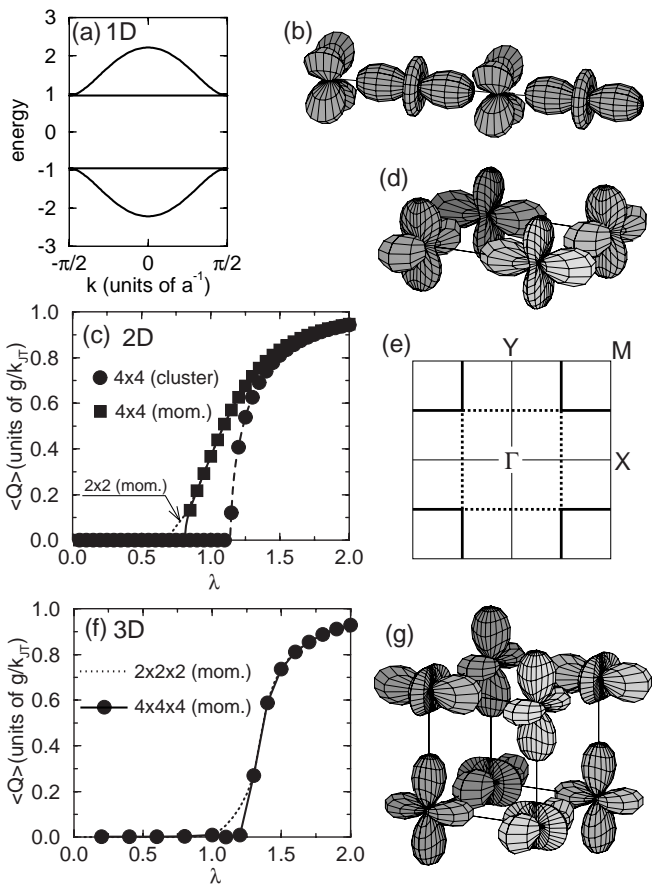}
\caption{(a) Energy band structure of the 1D chain at $\lambda$=1.
(b) Orbital ordering for the 1D case, with the alternation of
$3x^2$$-$$r^2$ and $y^2$$-$$z^2$.
(c) Average distortion as a function of $\lambda$ in the 2D plane.
Solid circles and squares denote full-optimized results for the 
4$\times$4 cluster and in momentum space, respectively,
while solid and dashed curves denote the results by using a single
parameter $q$ for the 4$\times$4 cluster and in momentum space, respectively.
Dotted curve indicates the full-optimized result in momentum
space using the 2$\times$2 lattice unit.
(d) Orbital ordering in the 2D plane, with the staggered pattern
of $Q_{2}$-type distortions.
(e) Fermi-surface lines of the 2D lattice. Solid and broken lines
indicate the Fermi-surface line for upper and lower band, respectively.
(f) Average distortion as a function of $\lambda$ for the 3D lattice.
Solid circles with curve 
denotes the full-optimized results in momentum space
by using 4$\times$4$\times$4 unit cell, while dotted curve indicates
those for the 2$\times$2$\times$ unit cell.
(g) Orbital ordering in the 3D lattice.
}
\end{figure}

In Fig.~1, the main results for $H_{\rm JT}$ are summarized.
For the 1D chain, it is emphasized that the JT distortion occurs
for {\it infinitesimal} values of $\lambda$, consistent
with Peierls instability concepts.
This issue can be clearly understood by the opening of the Peierls gap
in the band structure, as shown in Fig.~1(a).
In this case, the orbital ordering with the alternation of $3x^2$$-$$r^2$
and $y^2$$-$$z^2$ orbitals, as shown in Fig.~1(b),
is stabilized due to the cooperative JT distortions.\cite{Hotta3}
On the other hand, 
for the 2D case the metal-insulator transition can be observed
at a finite value of $\lambda$.
To observe this, it is convenient to monitor the average value of
the JT distortion
$\langle Q \rangle$=$(1/N)\sum_{\bf i} \sqrt{Q_{2{\bf i}}^2+Q_{3{\bf i}}^2}$,
where $N$ is the total number of sites.
As shown in Fig.~1(c) using a 4$\times$4 cluster, $\langle Q \rangle$ 
changes from zero to a finite value at $\lambda$$\sim$1.2,
a clear indication of a metal-insulator transition.
In this situation, the orbital-ordering pattern becomes bipartite
as shown in Fig.~1(d), indicating that
$Q_{2{\bf i}}$=$q$ for the ``A'' sublattice,
while $Q_{2{\bf i}}$=$-q$ for the ``B'' sublattice,
with $Q_{3{\bf i}}$=0 for all sites.
In other words, 
the distortion can be expressed by using a single parameter $q$,
since it occurs in a cooperative manner and all distortions
are correlated with one another.
Thus, our task can be drastically reduced in the numerical
optimization of distortions, since it is enough to minimize
the total energy as a function of $q$.
Such a calculation has been also performed and the result is
expressed as the dashed-curve in Fig.~1(c) for 4$\times$4 cluster,
in very good agreement with the results for the full optimization
within the computational error-bars.

To obtain the 4$\times$4 cluster result of
Fig.~1(c), anti-periodic boundary conditions are imposed
to satisfy the closed-shell condition for the electron configuration
in the non-interaction case. The small lattice size and use of special
boundary conditions may raise concerns about the existence of the
metal-insulator transition in the bulk limit.
To clarify lattice size effects,
calculations in momentum space
have been performed by imposing the twisted boundary condition
with the Bloch phase on the unit cell of 2$\times$2 or 4$\times$4 lattice.
The difference in the size of the unit cell appears in the sharpness
of the transition, indicating that the period of the distortion becomes
large at and near the transition region.
As shown by solid circles in Fig.~1(c), again we observe clear indications
of a metal-insulator transition, although the critical value is reduced.
This fact suggests that the present metal-insulator transition is
$not$ spurious of a finite-size cluster.
As mentioned above, in the 2D case, the distortion has been expressed
as a single parameter $q$.
This is also true in the results for the momentum space calculation,
as indicated by the solid curves in Fig.~1(c), again
in good agreement with the full-optimized results.

Let us consider the reason why the metallic phase can exist
even at half-filling. To clarify this point,
it is quite useful to depict the Fermi-surface.
As shown in Fig.~1(e), the nesting vector is $(\pi,0)$ or $(0,\pi)$,
{\it not} $(\pi,\pi)$. These nesting vectors are 
$not$ compatible with the 
staggered orbital ordering pattern (Fig.~1(d)) 
that is stabilized increasing $\lambda$. 
This is one of the remarkable features of the multiorbital
$e_{\rm g}$-electron system, which is not specific to two dimensionality.
In fact, in the results for the 3D case, as shown in Fig.~1(f),
we also observe the metal-insulator transition at a finite value of $\lambda$.
The dotted curves are obtained by using the 2$\times$2$\times$2 unit cell
where the twisted boundary condition with the Bloch phase is imposed,
while the solid circles are results for a
4$\times$4$\times$4 unit cell.
Except for the sharpness in the transition, in both cases, a
signal for a metal-insulator transition is clearly obtained.
In this case, the orbital ordering pattern becomes very complicated, as shown
in Fig.~1(g). Note that this pattern repeats periodically
on lattice larger than 2$\times$2$\times$2.
In the 3D case, an intrinsic incompatibility between the Fermi surface
and the orbital ordering pattern is also found.
Even without invoking the numerical results discussed before, 
the qualitative arguments presented here
related with the lack of nesting effects in 
$H_{\rm JT}$ at small $\lambda$ strongly suggests the presence of 
a metallic phase in two and three dimensions.

\begin{figure}[t]
\includegraphics[width=1.0\linewidth]{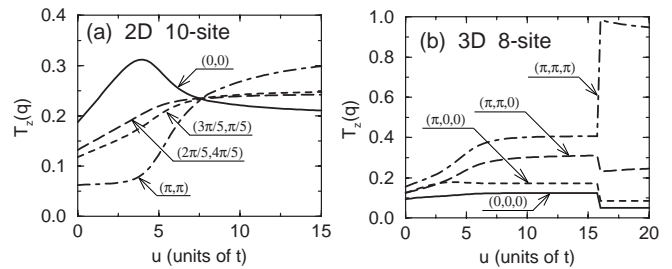}
\caption{Orbital correlation function $T_z({\bf q})$
as s function of $u$ (a) in the 2D plane with 10 sites
and (b) in the 3D cube with 8 sites.
}
\end{figure}


Now let us turn our attentions to the effect of Coulomb interactions
by considering the Hubbard-like model,
$H_{\rm C}$=$H_{\rm kin}$+$H_{\rm el-el}$,
namely dropping $H_{\rm el-ph}$ in this case.
This model is analyzed by exact diagonalization
to measure the orbital correlation defined as
$T_z({\bf q})$=$(1/N)$
$\sum_{\bf i,j} e^{i{\bf q}\cdot({\bf i}-{\bf j})}
\langle \tau_{z{\bf i}} \tau_{z{\bf j}} \rangle$.
In the 1D case (not shown), $T_z(\pi)$ becomes dominant for infinitesimal
value of $u$, indicating that the metallic phase does not exist
in the 1D systems except for the non-interacting case,
as is well known in the standard Hubbard model at half-filling.
However, in higher dimensions the situation is quite different.
Figure 2(a) denotes the results for a 2D lattice with 10 sites,
showing that the $(\pi,\pi)$ component becomes dominant for larger
value of $u$. 
However, as discussed above, the topology of the Fermi-surface 
is incompatible with the $(\pi,\pi)$ correlation, and as a consequence
$T_z(\pi,\pi)$
is no longer dominant in the small-$u$ region.\cite{comment}
In the 3D case with 2$\times$2$\times$2 cube, a sudden increase
of the $(\pi,\pi,\pi)$ component is observed at $u$$\sim$16,
since a level crossing occurs in this case.
The orbital ordering pattern shown in Fig.~1(g) does not show
$(\pi,\pi,\pi)$ correlation due to the cooperative effect.
Without the cooperative distortion, it is natural that $(\pi,\pi,\pi)$
correlation becomes dominant, as deduced from the large-$u$ limit.
Unfortunately, since it is difficult to enlarge the lattice size
for the exact diagonalization, it cannot be proven exactly that
the metal-insulator transition for $H_{\rm C}$ occurs in the bulk limit.
However, the results are highly suggestive that the staggered component
in the orbital correlation becomes dominant in the $e_{\rm g}$-orbital
Hubbard model only at large value of $u$. At small couplings, a metal
should be stabilized.

\begin{figure}[t]
\includegraphics[width=1.0\linewidth]{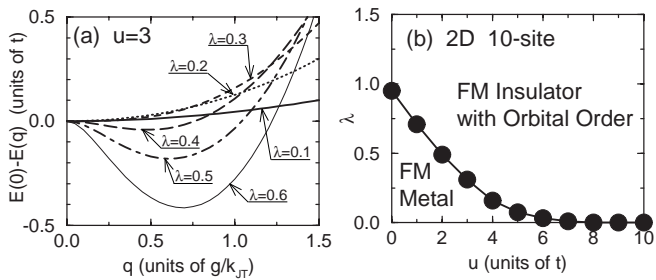}
\caption{(a) $E(0)$$-$$E(q)$ as a function of $q$ for $u$=3,
where $E(q)$ is the ground-state energy with lattice distortions
characterized by a single parameter $q$.
(b) Ground-state phase diagram in the $(\lambda,u)$-plane.
The boundary curve touches the $\lambda$=0 axis at $u$$\approx$7.5.
}
\end{figure}

Although it has been quite instructive to analyze the effects of
the electron-phonon coupling and Coulomb interaction
separately, both of them are simultaneously active
in real materials.
Thus, our final task here will be the study of the full model
$H$=$H_{\rm kin}$+$H_{\rm el-ph}$+$H_{\rm el-el}$
by a combination of
exact diagonalization for Coulomb interactions
and optimization techniques for lattice distortions.
It is a very hard task to perform the exact diagonalization
at each step during the relaxation of the lattice distortion,
but in order to depict the ground-state phase diagram,
it is sufficient to analyze how the total energy changes
when lattice deformations occur.
As mentioned in the above paragraph, fortunately in the 2D lattice
the cooperative distortion can be expressed using a single
parameter $q$.
In Fig.~3(a), the total energy difference is presented
as a function of $q$ for several values of $\lambda$ at $u$=3.
These results clearly indicate that the lattice distortion does not occur
for $\lambda$$\alt$0.3, while the optimized state accompanies the
lattice distortion for $\lambda$$\agt$0.3,
since the energy minimum is located at a finite value of $q$.
Thus, in order to detect the metal-insulator transition
under reasonable CPU times,
it is enough to monitor the change of the total energy for a very small
value of $q$.
If it is negative, it strongly suggests the occurence of lattice distortions.
In actual calculations, the energy difference is evaluated for
$q$=0.01. Using this procedure,
 the phase diagram for the 2D 10-site cluster is shown in Fig.~3(b).
The results clearly indicate
that the metallic phase exists, 
even in this case where 
both electron-phonon and Coulomb interactions are included.
Note that the transition at $\lambda$=0
appears as a cross-over behavior as indicated above,
but in the present calculations for non-zero $\lambda$,
the boundary curve seems to converge to the cross-over point
at $u$$\approx$7.5, consistent with Fig.~2(a).

Finally, the novel possibility of observing CMR effects in undoped manganites
is here discussed.
As mentioned in the introductory paragraph, in the theoretical phase diagram
for the undoped limit, the FM metallic state has been observed adjacent to
the A- or E-type AFM insulating phase.\cite{Hotta2}
In the present work, it has been confirmed that such a metallic phase is
$not$ spuriously caused by the small-cluster calculations,
and moreover the phase  does not disappear by adding Coulomb interactions.
Thus, within the theoretical framework for CMR that it is based on
the competition between FM metallic and insulating phases of any nature,  
{\it CMR-like phenomena could occur even in undoped manganites},
a very unexpected and challenging theoretical prediction.


In summary, the metal-insulator transition for undoped manganites
has been discussed based on the $e_{\rm g}$ orbital Hubbard model
tightly coupled to JT distortions.
The existence of a novel FM metallic phase at zero hole doping has been
shown, leading to the intriguing
possibility of CMR effects even in  RMnO$_3$.


The author thanks E. Dagotto and K. Ueda for fruitful discussions
and valuable comments.
He is supported by the Grant-in-Aid for Scientific
Research from Japan Society for the Promotion of Science.



\begin{references}

\bibitem{Tokura}
See, for instance, {\it Colossal Magnetoresistance Oxides},
edited by Y. Tokura, Gordon \& Breach, New York, 2000.

\bibitem{Dagotto1}
E. Dagotto {\it et al.}, Phys. Rep. {\bf 344}, 1 (2001).

\bibitem{Dagotto2}
E. Dagotto,
{\it Nanoscale Phase Separation and Colossal Magnetoresistance},
Springer-Verlag, Berlin, 2002.

\bibitem{Hotta1}
T. Hotta {\it et al.}, Phys. Rev. B{\bf 60}, R15009 (1999)
and references for LaMnO$_3$ therein.

\bibitem{Munoz}
A. Mu\~noz {\it et al.}, Inorg. Chem. {\bf 40}, 1020 (2001).

\bibitem{Kimura}
T. Kimura {\it et al.}, cond-mat/0211568.

\bibitem{Hotta2}
T. Hotta {\it et al.}, cond-mat/0211049.

\bibitem{Yunoki}
S. Yunoki {\it et al.}, Phys. Rev. Lett. {\bf 81}, 5612 (1998).

\bibitem{Hotta3}
T. Hotta {\it et al.}, Phys. Rev. B{\bf 62}, 9432 (2000).

\bibitem{note}
Effect of the breathing-mode distortion is not important as long as $\beta$
is larger than unity, as discussed previously in Ref.~\onlinecite{Hotta1}.
In actual manganites, $\beta$ is estimated to be about two, indicating that
$Q_{1{\bf i}}$ does not play an active role. Throughout this paper,
$\beta$ is fixed as two or larger, without further remarks.

\bibitem{comment}
The peculiar increase in $T_z(0,0)$ in the small-$u$ region may be due to
the special properties of the 10-site cluster, 
since it is not found in the 8-site result.

\end{references}
\end{document}